\documentclass{appolb}
\usepackage{epsfig}
\newcommand{\lsim}{\raisebox{-0.13cm}{~\shortstack{$<$ \\[-0.07cm] $\sim$}}~}

\newcommand{\tb}{\tan \beta}

\newcommand{\beq}{\begin{eqnarray}}
\newcommand{\eeq}{\end{eqnarray}}
\newcommand{\mchi}{{m_{\chi_1^0}}}
\newcommand{\lsp}{{\chi_1^0}}

\begin{document}
\title{Dark Matter and the ILC%
\thanks{Presented at  PHOTON2005, Warsaw, 31/08--04/09.}%
}
\author{Abdelhak DJOUADI
\address{LPT, Universit\'e Paris--Sud, F--91405 Orsay, France}
}
\maketitle
\begin{abstract}
We discuss the solution to the Dark Matter problem provided by the lightest
neutralino of the Minimal Supersymmetric Standard Model (MSSM) and highlight 
the role of the International Linear Collider (ILC) in determining its
cosmological relic density.  
\end{abstract}
  
\section{Introduction}

As deduced from the WMAP satellite measurement of the temperature anisotropies
in the Cosmic Microwave Background, in combination with data on the Hubble
expansion and the density fluctuations in the universe, cold Dark Matter (DM) 
makes up $\approx 25\%$ of the energy of the universe~\cite{WMAP}. The DM
cosmological relic density is precisely measured to be 
\beq
\Omega_{\rm DM}\, h^2 = 0.113 \pm 0.009 
\label{omrange}
\eeq
which leads to $0.087 \leq \Omega_{\rm DM}\, h^2 \leq 0.138$ at the 99\%
confidence level. In these equations, $\Omega \equiv \rho / \rho_c$, where
$\rho_c \simeq 2 \cdot 10^{-29} h^2 {\rm g/cm^3}$ is the ``critical'' mass
density that yields a flat universe, as favored by inflationary cosmology and
as verified by the WMAP satellite itself; $\rho < \rho_c$ and $\rho >\rho_c$
correspond, respectively to an open and closed universe, i.e. a metric with
negative or positive curvature. The dimensionless parameter $h$ is the scaled
Hubble constant describing the expansion of the universe.

In the Minimal Supersymmetric Standard Model (MSSM), there is an ideal 
candidate for the weakly interacting massive particle (WIMP) which is expected 
to form this cold DM: the lightest neutralino $\chi_1^0$ which is a mixture of
the supersymmetric partners of the neutral gauge and Higgs bosons and is in
general the Lightest Supersymmetric Particle (LSP). This electrically neutral
particle is  absolutely stable when a symmetry called R--parity is conserved,
is  massive and thus non--relativistic or cold. Furthermore, it has only weak
interactions and for a wide range of the MSSM parameter space, its annihilation
rate into SM particles fulfills the requirement that the resulting cosmological
relic density is within the range measured by WMAP. This is particularly the
case in the widely studied minimal Supergravity (mSUGRA)  scenario \cite{nilles}
and in some of its  variants; see Ref. \cite{Reviews}.

In this brief note, we  discuss the contribution of the LSP neutralino to
the overall matter density of the universe and highlight the role of the 
International Linear Collider (ILC)  in determining this relic density.
 
\section{The Dark Matter relic density}

To derive the cosmological relic density, the standard treatment \cite{DM} is
based on  the assumption [besides that the LSP should be effectively stable,
i.e. its lifetime should be long compared to the age of the Universe, which
holds in the MSSM with conserved R--parity that is discussed here] that the
temperature of the Universe after the last period of entropy production must
exceed $\sim 10\%$ of $\mchi$, an assumption which is quite natural in the
framework of inflationary models \cite{DM}.  In the early universe all
particles were abundantly produced and were in thermal equilibrium through
annihilation and production processes. The time evolution of the number density
of the particles is governed by the Boltzmann equation
\beq 
\frac{d n_\lsp} {dt} +3H n_\lsp =- \langle v\, \sigma_{\rm ann}  \rangle 
( n_\lsp^2- n_\lsp^{\rm eq\,2}) 
\eeq 
where $v$ is the relative LSP velocity in their center--of--mass frame,
$\sigma_{\rm ann}$ is the LSP annihilation cross section into SM particles and
$\langle \dots \rangle$ denotes thermal averaging; $n_\lsp$ is the actual
number density, while $n_\lsp^{\rm eq}$ is the thermal equilibrium number
density. The Hubble term takes care of the decrease in number density due to
the expansion, while the first and second terms on the right hand side
represent, respectively,  the decrease due to annihilation and the increase
through creation by the inverse reactions. If the assumptions mentioned above 
hold, $\chi_1^0$ decouples from the thermal bath of SM particles at an inverse
scaled temperature $x_F \equiv \mchi / T_F$ which is given by \cite{Reviews}
\beq \label{edm1}
x_F = 0.38 M_{P} \langle v \sigma_{\rm ann} \rangle c (c+2)
m_{\chi_1^0}  \, (g_* x_F )^{-1/2}
\eeq
where $M_{P}\!=\!2.4\!\cdot\!10^{18}$ GeV is the (reduced) Planck mass, $g_*$ 
the number of relativistic degrees of freedom which is typically $g_* \simeq
80$ at $T_F$, and $c$ a numerical constant which is taken to be $\frac12$;  
one typically finds $x_F \simeq 20$ to 25. Today's LSP density in units of the 
critical density is then given by \cite{Reviews}
\beq \label{edm2}
\Omega_\chi h^2 = \frac {2.13 \cdot 10^8 / {\rm GeV}} { \sqrt{g_*} M_{P} 
J(x_F)} \, , \ {\rm with} \ \ J(x_F) = \int_{x_F}^\infty \frac {\langle v 
\sigma_{\rm ann} \rangle(x) } {x^2} dx
\eeq
Eqs.~(\ref{edm1})--(\ref{edm2}) provide an approximate solution of the 
Boltzmann equation which has been shown to describe the exact numerical 
solution very accurately for all known scenarios.
Since $\chi_1^0$ decouples at a temperature $T_F \ll m_\chi$, in most cases it
is sufficient to use an expansion of the LSP annihilation rate in powers of the
relative velocity between the LSPs
\beq \label{edm4}
v \, \sigma_{\rm ann} \equiv v \, \sigma(\chi_1^0 \chi_1^0 \rightarrow {\rm
SM\, particles}) = a + b v^2 + {\cal O} (v^4)
\eeq
The entire dependence on the model parameters  is then contained in the
coefficients $a$ and $b$, which essentially describe the LSP annihilation cross
section from an initial S-- and P--wave, since the expansion of the
annihilation rate of eq.~(\ref{edm4}) is only up to ${\cal O}(v^2) $.
S--wave contributions start at ${\cal O}(1)$ and contain ${\cal O}(v^2)$
terms that contribute to eq.~(\ref{edm4}) via interference with the ${\cal O}
(1)$ terms. In contrast, P--wave matrix elements start at ${\cal O}(v)$, so
that only the leading term in the expansion is needed. There is no interference
between S-- and P--wave contributions, and hence no ${\cal O}(v)$ terms.  

In generic scenarios the expansion eq.~(\ref{edm4}) reproduces exact results 
quite well. However, it fails in some exceptional cases \cite{Reviews} all of
which can be realized in some part of the MSSM parameter space, and even in
mSUGRA: 

$i$) The expansion breaks down near the threshold for the production of
heavy particles, where the cross section depends very sensitively on the c.m.
energy $\sqrt{s}$. In particular, due to the non--vanishing kinetic energy of
the neutralinos, annihilation into final states with mass exceeding twice the
LSP mass (``sub--threshold annihilation'') is possible. This is particularly
important in the case of neutralino annihilation into $W^+W^-$ and $hh$ pairs, 
for relatively light higgsino--like and mixed LSPs, respectively.

$ii$) The expansion eq.~(\ref{edm4}) also fails near $s-$channel poles,
where the cross section again varies rapidly with $\sqrt{s}$. In the MSSM, this
happens if twice the LSP mass is near $M_Z$, or near the mass of one of the
neutral Higgs bosons.  In models with universal gaugino
masses, the $Z$-- pole region is now excluded by chargino searches at LEP2
and we are left only with the Higgs pole regions which are important as will be
seen later.

$iii$) If the mass splitting between the LSP and the next--to--lightest
superparticle NLSP is less than a few times $T_F$, co--annihilation processes
involving one LSP and one NLSP, or two NLSPs, can be important
\cite{co-annihilation}.  Co--annihilation is important in three cases: higgsino
or SU(2) gaugino like LSPs  and when the LSP is degenerate in  mass with
$\tilde{\tau}_1$  or with the lightest top squark (the latter case hardly occurs
in mSUGRA scenarios).

\section{The relic density in the mSUGRA scenario}

The mSUGRA model \cite{nilles} is the most widely studied implementation
of the MSSM and it manages to describe phenomenologically acceptable spectra 
with only four parameters plus a sign:
\beq \label{paras}
m_0, \ m_{1/2}, \ A_0, \ \tb, \ {\rm sign}{\mu}.
\eeq
where $m_0,m_{1/2}$ and $A_0$ are the common soft SUSY--breaking terms of all
scalar masses, gaugino masses and trilinear scalar interactions, defined at
at the Grand Unification scale. $\tb$ is the ratio of the vacuum expectation 
values (vev's) of the two Higgs doublets at the weak scale and $\mu$ is
the supersymmetric higgs(ino) mass parameter.

We use the Fortran code SUSPECT \cite{suspect} to solve the RGE and to
calculate the spectrum of physical sparticles and Higgs bosons, following the
procedure outlined in Ref.~\cite{ddk}. In addition to leading to consistent
electroweak symmetry breaking (EWSB), a given set of input parameters has to
satisfy experimental constraints \cite{pdg}. The ones relevant for this study
are \cite{ddk}:

-- The total cross section for the production of any pair of sparticles
  at the highest LEP energy (209 GeV) must be less than 20 fb. 
  
-- Searches for neutral Higgs bosons at LEP  impose a lower bound on $m_h$;
  allowing for a theoretical uncertainty, one  requires $m_h>$  111 GeV.

-- Recent measurements  of the muon magnetic moment  lead to the constraint on
  the SUSY contribution: $ -5.7 \cdot 10^{-10} \leq a_{\mu,\, {\rm SUSY}} \leq
  4.7 \cdot 10^{-9}$.

--   Allowing for experimental and theoretical errors, the
  branching ratio for radiative $b$ decays should be $
  2.65 \cdot 10^{-4} \leq B(b \rightarrow s \gamma) \leq 4.45 \cdot 10^{-4}$. 
  
-- Finally, the calculated $\tilde \chi_1^0$ relic density has to be in the range
  (\ref{omrange}).

\begin{figure}[h!]
${\mathbf m_0}$
\vspace*{-.3cm}
\begin{center}
\mbox{\epsfig{file=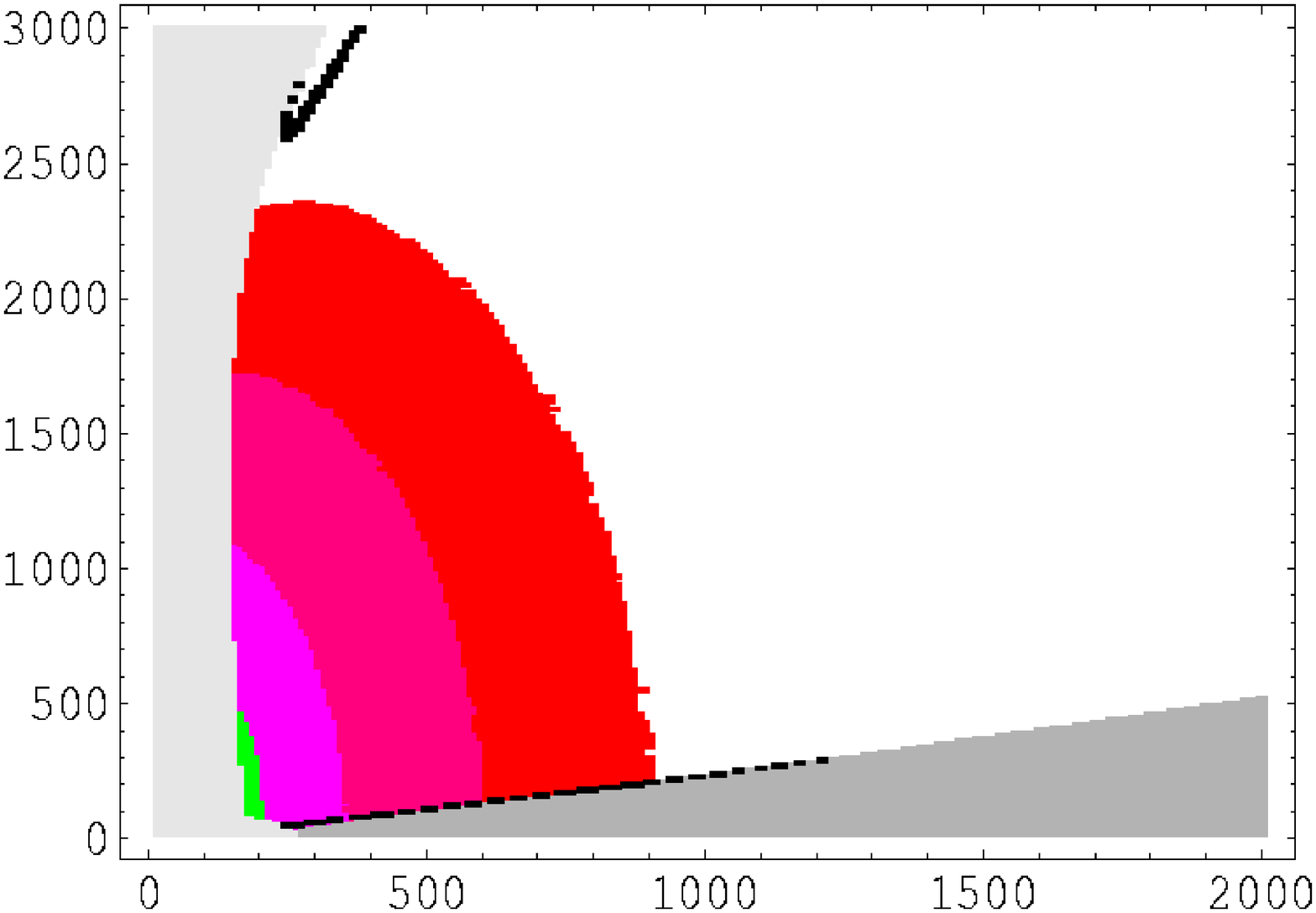,width=5.5cm,height=5.5cm}\hspace*{0.9cm}
      \epsfig{file=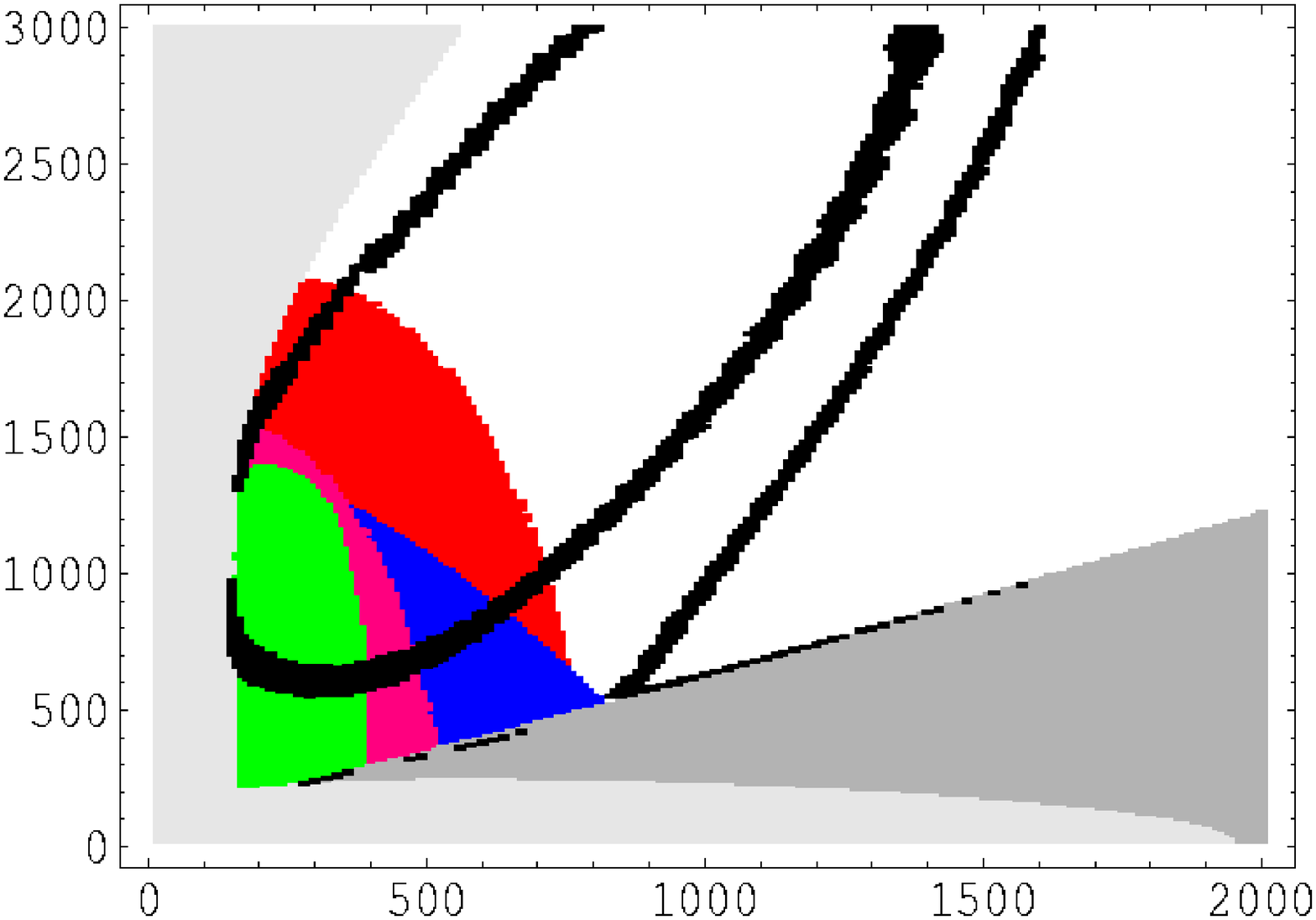,width=5.5cm,height=5.5cm} }
\hspace*{11.3cm} ${\mathbf m_{1/2}}$
\vspace*{-.3cm}
\end{center}

\caption{\it The mSUGRA $(m_{1/2},m_0)$ parameter space with all constraints 
  imposed for $A_0=0, \mu>0$ and $\tan\beta=10$ (left) and 50 (right). The top
  quark mass is fixed to the new central value, $m_t=172.7$ GeV. The light grey
  regions are excluded by the requirement of correct electroweak symmetry
  breaking, or by sparticle search limits. In the dark grey regions
  $\tilde\tau_1$ would be the LSP. The light pink regions are excluded by
  searches for neutral Higgs bosons at LEP, whereas the green regions are
  excluded by the $b \rightarrow s \gamma$ constraint. In the blue
  region, the SUSY contribution to $g_\mu-2$ falls in the correct range 
   whereas the red regions are compatible with having an SM--like Higgs boson
  near 115 GeV. Finally, the black regions satisfy the DM constraint.}
\vspace*{-.6cm}
\end{figure}

The output \cite{ddk} is shown and partly commented in Fig.1. The black regions are those
satisfying the DM constraint, eq.~(1). In general, there are four familiar 
regions where this constraint is satisfied. $i)$ Scenarios where both $m_0$ and
$m_{1/2}$ are rather small (the ``bulk region'') are most natural from the point
of view of EWSB but are severely squeezed by lower bounds from searches for
sparticles and Higgs bosons. $ii)$ In the ``co--annihilation'' region one has
$\mchi \simeq m_{\tilde \tau_1}$, leading to enhanced destruction of sparticles
since the $\tilde \tau_1$ annihilation cross section is about ten times larger
than that of the LSP; this requires $m_{1/2} \gg m_0$. $iii)$ The ``focus
point'' or ``hyperbolical branch'' region occurs at $m_0 \gg m_{1/2}$, and
allows $\tilde \chi_1^0$ to have a significant higgsino component, enhancing its
annihilation cross sections into final states containing gauge and/or Higgs
bosons; however, if $m_t$ is much higher than its current central value  of 173
GeV, this solution requires multi--TeV scalar masses. $iv)$ Finally, if
$\tan\beta$ is large, the $s-$channel exchange of the CP--odd Higgs boson $A$
can become nearly resonant, again leading to an acceptable relic density (the
``$A-$pole'' region).

Recently, a fifth cosmologically acceptable region of mSUGRA parameter space has
been revived \cite{hpole}. In a significant region of parameter space one has $2
\mchi \lsim m_h$, so that $s-$channel $h$ exchange is nearly resonant. This
``$h-$pole'' region featured prominently in early discussions of the DM density
in mSUGRA  but seemed to be all but excluded by the combination of rising lower
bounds on $m_h$ and $\mchi$ from searches at LEP \cite{pdg}. However, in recent
years improved calculations \cite{newhiggs} of the mass of the light CP--even
$h$ boson have resurrected this possibility for top mass values close to or
larger than 178 GeV (thus, this region does not appear in Fig.~1).

\section{SUSY particle masses and the ILC}

Bounds on physical masses might be a more meaningful way to show the
possibilities of mSUGRA than the ubiquitous plots of allowed regions in the
space of basic input parameters, Fig.~1, in which one  always fixes the values
of some other free parameters (e.g., $A_0$, $\tb$, $m_t$). One obtains the least
biased view of the allowed ranges of masses by simply scanning over the entire
parameter set that is consistent with a given set of constraints. 

\begin{table}[h!]
\vspace*{-1mm}
\begin{center}
\begin{tabular}{|c||c|c|c|}
\hline
  (s)particle
  & \multicolumn{3}{c|}{mass bounds [GeV]} \\
 & \ \ \ \ \ \ \ \ Set I\ \ \ \ & \ \ \ \ Set II\ \ \ \  & \ \ \ \ \ Set III\
  \ \ \ \  \\
\hline
$\tilde \chi_1^0$ & 50 & 53 & [53, 61] \\
$\tilde \chi_1^\pm$ & 105 & 105 & [105, 122] \\
$\tilde \chi_3^0$ & 136 & 137 & [280, --] \\
\hline
$\tilde \tau_1$ & 99 & 99 & [630, --] \\
$h$ & 114 & 114 & [114, 122] \\
$H^\pm$ & 128 & 128 & [246, --] \\
\hline
$\tilde g$ & 374 & 383 & [383, 482] \\
$\tilde d_R$ & 444 & 444 & [774, --] \\
$\tilde t_1$ & 102 & 110 & [110, --] \\
\hline
\end{tabular}
\vspace*{2mm}
\caption{\it Sparticle mass bounds in mSUGRA obtained by scanning over the entire
  allowed parameter space, defined by $m_t \in [171 \, {\rm GeV}, \, 185 \, {\rm
  GeV}], \ (m_{\tilde t_1} + m_{\tilde t_2})/2 \leq 2$ TeV, the lower bounds on
  sparticle and Higgs masses from collider experiments, the constraint on $b
  \rightarrow s \gamma$, simple `CCB'  constraints  and a conservative
  interpretation of the constraint from  $g_\mu-2$ (essentially the overlap of
  the $2\sigma$ regions using $\tau$ decay and $e^+e^-$ collider data). Set II
  adds the DM constraint to the above set of constraints. Set III is like Set
  II, except that the scanned region has been artifically limited to the $h-$pole
  region, where $m_{\tilde \chi_1^0} \leq m_h/2$. Only {\em lower} bounds are
  listed for Sets I and II, while for Set III the allowed range is given; a dash
  (--) means that the upper bound is directly set by the upper bound
  on the average stop mass.}
\label{tab:a}
\vspace*{-5mm}
\end{center}
\end{table}

Table 1 lists lower bounds on the masses of some new (s)particles in mSUGRA,
first without (set I) and then with (set II)  the DM constraint. The lower
bounds on many new (s)particles simply coincide with the bounds established by
collider experiments. This is true for the lighter chargino, stau and scalar
Higgs states, and essentially also holds true for the lighter stop. The bounds
on the masses of the gluino and third neutralino are essentially the same as
that in a more general MSSM, as long as gaugino mass unification is maintained.
Clearly the DM constraint still allows some new (s)particles to be quite light.
One should emphasize, however, that usually the lower bounds in the table cannot
be saturated simultaneously.  Nevertheless, the possibility of light sparticles
even in this simplest of all potentially realistic SUSY models that allow WIMP
Dark Matter should be quite encouraging to experiments.

Set III shows these bounds (including the DM constraint) when one confines
oneself to the $h-$pole region discussed at the end of the previous subsection.
In this case there are significant upper bounds on the masses of all gauginos. 
The reason is that one needs $2 m_{\tilde \chi_1^0} \simeq m_h \leq 120$ GeV
here, leading in mSUGRA with the assumed universality of gaugino masses at the GUT
scale, to relatively light charginos and neutralinos as well as gluinos. 

\section{The determination of the relic density at the ILC}

Thus, in many scenarios SUSY particles can be produced abundantly at the  next
generation of high--energy colliders, in particular at the LHC and the ILC.   
However, to determine the predicted WIMP relic density (see the flowchart shown
in the left-hand side of Fig.~2), one must experimentally constrain all
processes contributing  to the LSP pair annihilation cross section;  this
requires detailed knowledge not only of the LSP properties, but also of all
other particles contributing to their annihilation.  This is not a simple task
and all unknown parameters entering the determination of $\Omega_ \chi h^2$ need
to be experimentally measured or shown to have marginal effects. 

 Many high precision measurements are possible at the LHC \cite{LHC-DM}, but
many of them can be vastly improved at the ILC \cite{ILC-DM,Feng}. Because of
the clean environment and the knowledge of the c.m. energy of the initial beams,
sparticle masses can be determined with high accuracy through kinematic
endpoints and threshold scans. The results of  one study \cite{Feng} in a given
mSUGRA scenario (SPS1a point \cite{SPS})  are summarized in the  right-hand
side  of Fig.~2, where the achievable precision at collider experiments are
compared with the satellite determination of $\Omega_ \chi h^2$.  The figure
shows that the ILC will provide a part per mille determination of $\Omega_\chi
h^2$ in the case under study, matching WMAP and even the huge accuracy expected
from Planck.  The many possible implications of such measurements are outlined
in the flowchart in the right--hand side  of the figure.

\begin{figure}[htbp]
\begin{center}
\mbox{\includegraphics[width=0.5\textwidth]{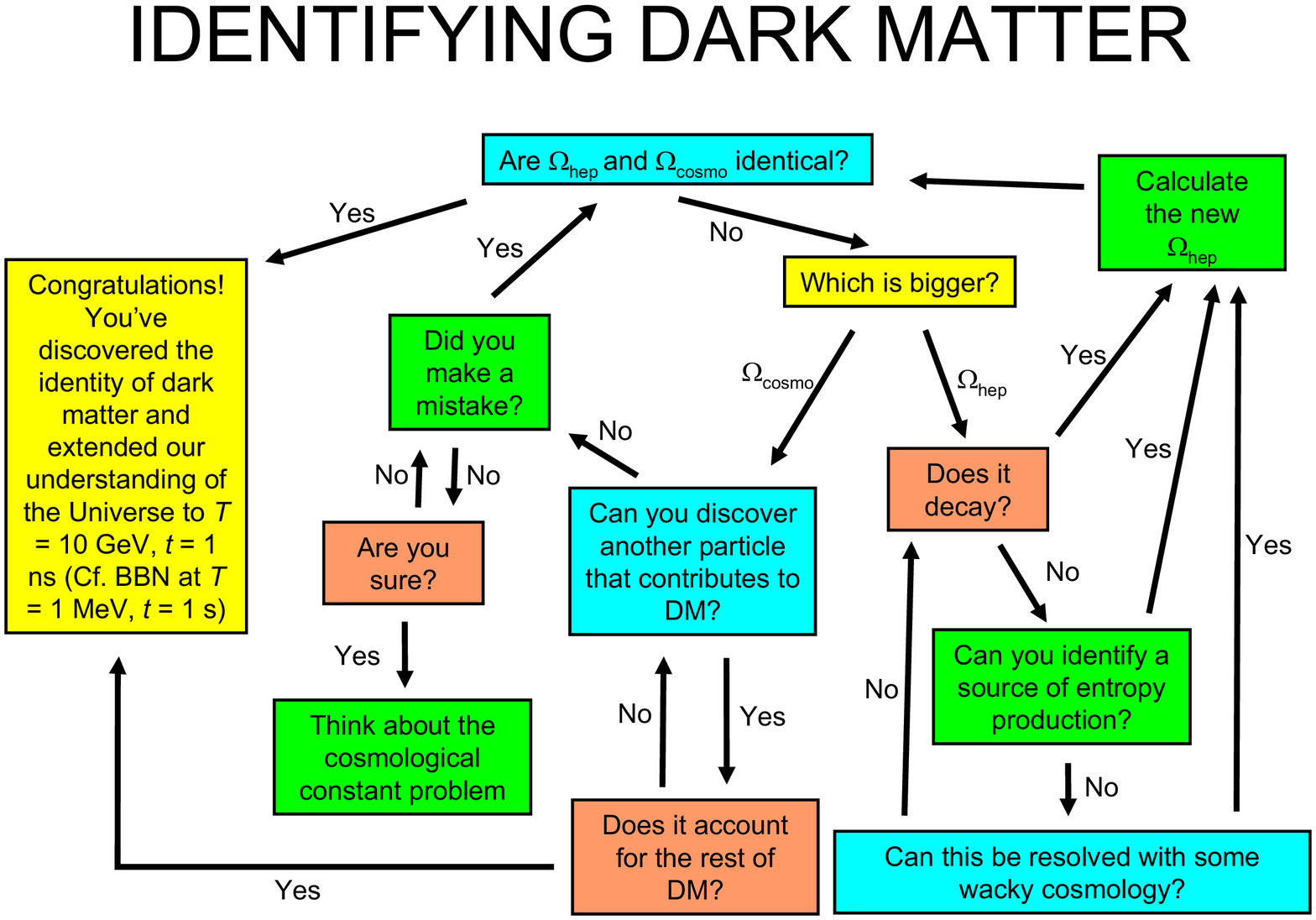} 
\includegraphics[width=0.5\textwidth]{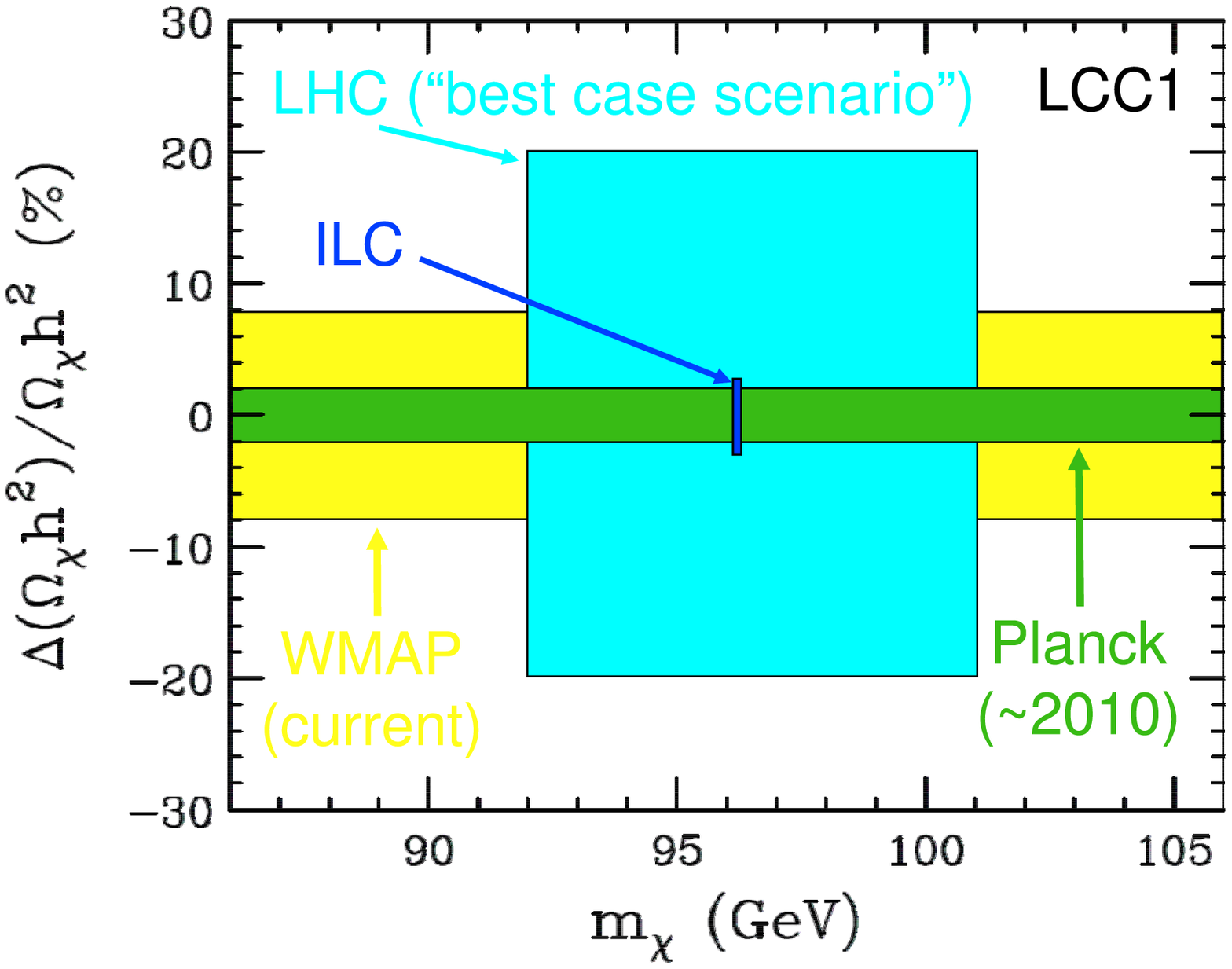}
}
\end{center}
\caption{\it Left: Flowchart illustrating the possible implications of 
comparing $\Omega_{hep}$, the predicted DM density determined from  high
energy physics, and $\Omega_{cosmo}$, the actual DM density determined by
WMAP and Planck. Right: Constraints in the $(\mchi, \Delta (\Omega_\chi h^2)/
\Omega_\chi h^2 )$ plane from the ILC and LHC.  Constraints on $\Delta 
(\Omega_\chi h^2)/ \Omega_\chi h^2$ from WMAP and Planck (which provide no 
constraints on  $\mchi$) are also shown; from Ref.~\cite{Feng}.}
\end{figure}

Thus, if DM is composed of the lightest neutralinos, the LHC and particularly
the ILC will be able to determine the WIMP's properties and  pin down its relic
density.  If these determinations match cosmological observations to high
precision, then (and only then) we will be able to claim to have determined what
dark matter is. Such an achievement would be a great success of the
particle physics/cosmology connection and would give us confidence in
our understanding of the Universe.\bigskip

{\bf Acknowledgements:} I would like to thank the organizers of this very
successful  conference, in particular Maria Krawczyk, for their invitation and
for the very nice and stimulating atmosphere. Thanks also to Manuel Drees and
Jean Loic Kneur, for the enjoyable collaboration which led to some of the
results discussed here.

\end{document}